\documentclass{ws-mplb}

\begin{document}

\markboth{A. L. de Paula Jr and D. S. Freitas}{Dynamics of entanglement among the environment oscillators of a many-body systems}

%
\catchline{}{}{}{}{}
%

\title{DYNAMICS OF ENTANGLEMENT AMONG THE ENVIRONMENT OSCILLATORS OF A MANY-BODY SYSTEMS}

\author{A. L. de PAULA JR}

\address{Departamento de F\'\i sica,
Instituto de Ci\^encias Exatas, Universidade Federal de Minas
Gerais, 30123-970, Belo Horizonte, Minas Gerais,
Brazil\\
albertofisica@hotmail.com}

\author{DAGOBERTO S. FREITAS}

\address{Departamento de F\'\i sica, Universidade
Estadual de Feira de Santana, 44036-900, Feira de Santana, Bahia,
Brazil\\
dfreitas@uefs.br}

\maketitle

\begin{history}
\received{(Day Month Year)}
\revised{(Day Month Year)}
\end{history}

\begin{abstract}
In this work we extend the discussion that was began in the Ref. \cite{alberto14} [A. L. de Paula Jr., J. G. G. de Oliveira Jr., J. G. Peixoto de Faria,
Dagoberto S. Freitas and M. C. Nemes, {\it Phys. Rev.} {\bf A89}, (2014) 022303] to deal with the dynamics of the concurrence of a many-body system. In that previous paper, the discussion was focused in the residual entanglement between the partitions of the system. The purpose of the present contribution is to shed some light on the dynamical properties of entanglement among the environment oscillators. We consider a system consisting of a harmonic oscillator linearly coupled to N others and solve the corresponding dynamical problem analytically. We divide the environment in arbitrary two partitions and the entanglement dynamics between any of these partitions is quantified and it shows that in the case that excitations in each partitions are equal, the concurrence reaches the value one and the two partition of the environment are maximally entangled. For long times the excitations of the main oscillator are completely transferred to environment and the environment oscillators are found entangled.
\end{abstract}


\section{Introduction}
Entanglement, a property at the heart of Quantum Mechanics, has first been brought to scientific debate by the intriguing questions
posed by Einstein, Podolsky, and Rosen \cite{epr35} and since then the matter has always been under investigation. Several years ago,
the interest of the physical community in this counterintuitive property has been raised even more due to its potential as a resource for
information processing and quantum computation \cite{nielsen00}. The growing interest in the area of quantum information opened questions
also in mathematics as well as propelled the development of high precision experiments in several areas of physics (condensed matter,
quantum optics, and atomic physics) just to quote a few. In order to realize the goal of quantum computation several important issues
must be addressed. It thus has become clear that the search for entangled states or classes of entangled states constitutes an
important subject. Clearly understanding systems involving many entangled degrees of freedom is also an important part of quantum
computation. Therefore, much work has been devoted to the subject in the recent literature \cite{mohamed04,osborne06,kim09,chan10,mckeown10,li10,marcus11,ma11,chen12,osterloh13}. In \cite{mohamed04} is presented a scheme to experimental detection of genuine multipartite entanglement using
entanglement witness operators.  This method was applied to three and four-qubit entangled states of polarized photons, giving
experimental evidence that the considered states contain true multipartite entanglement. Multipartite qubit systems have also
been considered in \cite{osborne06,kim09,chan10}. Relevant new features appearing in multipartite systems as compared to bipartite
measures are highlighted together with an outline of how to experimentally measure multipartite entanglement \cite{mckeown10,li10,marcus11}.
A notion of generalized concurrence called genuine multipartite entanglement (GME) concurrence \cite{ma11,chen12} was introduced in the attempt
to distinguish between GME and partial entanglement and it was pointed out that GME concurrence of pure states may be directly accessible in laboratory
experiments. Many-particle systems, as far as their entanglement, are presented in Ref. \cite{osterloh13} and a brief
review of the bipartite measures of entanglement and the entanglement of pairs both for systems of distinguishable and indistinguishable
particles are shown. In this reference the concept of genuine multipartite entanglement is introduced and an approach for the
construction of genuine multipartite entanglement measures is presented. Solid mathematical results regarding multipartite
entanglement are only available for pure three-qubit systems \cite{coffman00}. In this case the existence of what is called
residual entanglement, not detectable by two-qubit entanglement quantifiers, has been proved \cite{hill97,wootters98}. To the residual
entanglement, in Ref. \cite{alberto14}, was showed that in any arbitrary three partitions of a many-body system a residual entanglement is always
present. In this context there appears two very distinct regimes governed by the overlap between the states in the superposition.

In this work we will extend the discussion that was began in the Ref. \cite{alberto14} to deal with the dynamics of the concurrence of a
many-body system. In that previous paper, the discussion was focused in the residual entanglement between the partitions of the system.
Similar to Ref. \cite{alberto14}, the purpose of the present contribution is to shed some light on the dynamical properties of
entanglement. In this paper, we will study the
dynamics among the environment oscillators. We will show that even without a direct interaction among the environment oscillators, they entangle with one another.

\section{The model}
\begin{figure}[th]
\centerline{\psfig{file=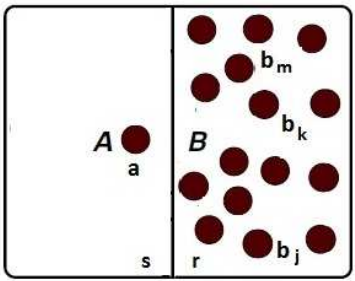,width=6cm}}
\vspace*{8pt}
\caption{Model for a harmonic oscillator  linearly coupled to $N$ other oscillators. The  circle in $A$ side represents the main oscillator, which is coupled to a large number of the environment oscillator ($B$ side ) with different frequencies $\omega_k$ via the coupling constants $\gamma_k$.\label{fig1}}
\end{figure}

We consider a system constituted by a harmonic oscillator linearly coupled to $N$ other oscillators; see Fig. (\ref{fig1}). The
Hamiltonian is given by
\begin{equation}
H=\hbar\omega_0a^{\dag}a +
\sum_{k}\hbar\omega_kb^{\dag}_{k}b_{k}+\sum_{k}\hbar\gamma_k\bigl(a^\dag
b_k+ab^{\dag}_{k}\bigr), \label{hamiltoniano}
\end{equation}
where the operators $a$ ($a^{\dag}$) and $b_k$ ($b_{k}^{\dag}$) correspond to annihilation (creation) operators of the main oscillator
and of the environment oscillator, respectively. $\omega_0$, $\omega_k$ correspond to the frequency of the main oscillator and of the environment oscillator. The symbol $\gamma_k$ is a coupling constant between the main oscillator and the $kth$ oscillator of the environment
with  $k=1,2,3,\ldots,N$.
\section{Dynamics}

\subsection{Dynamics of the system}
As the first step to solve exactly the Hamiltonian above, in higher-dimensional quantum systems, we will consider the initial
state of the main oscillator in a coherent state $|\alpha(0)\rangle$ and each given oscillator of the environment in the vacuum state
$|0_k(0)\rangle$. The initial state of the global system in this manner is given by
\begin{equation}
|\psi(0)\rangle=|\alpha(0)\rangle\prod_k|0_k(0)\rangle.\label{estado_inicial_fatorado}
\end{equation}
For any time later the state of the system has the form
\begin{equation}
|\psi(t)\rangle=|\alpha(t)\rangle\prod_k|\lambda_k(t)\rangle,
\label{estado_final_fatorado}
\end{equation}
where $|\alpha(t)\rangle$ and $|\lambda_k(t)\rangle$ are coherent states for any time later of the main oscillator and of the environment
respectively. The dynamic of the amplitudes $\alpha(t)$ and $\lambda_k(t)$ can be obtained by using (\ref{estado_final_fatorado}) in
the Schr\"odinger equation
\begin{equation}
i\hbar\frac{d}{dt}|\psi(t)\rangle = H|\psi(t)\rangle.\label{eqsch}
\end{equation}
By using the equation given by Eq. (\ref{estado_final_fatorado}) and developing the left side of the Schr\"odinger equation we obtain the
following set of equations:
\begin{eqnarray}
i\frac{d}{dt}f(t) &=& \sum_k \gamma_k g_k(t),\label{dif_f_01}\\
i\frac{d}{dt}g_k(t) &=& -2\delta_k g_k(t) +\gamma_k f(t),\label{dif_g_01}
\end{eqnarray}
where we express $\alpha(t)$ and $\lambda_k(t)$ in the form
\begin{eqnarray}
\alpha(t) &=& \alpha(0) e^{-i \omega_0 t}f(t),\label{alpha_01}\\
\lambda_k(t) &=&\alpha(0) e^{-i \omega_k t}g_k(t)e^{-2i\delta_kt}.\label{lambda_01}
\end{eqnarray}
In the set of equations above $f(t)$ and $g_{k}(t)$ are amplitudes that contain information
about the number of the excitations in each subsystems and $\delta_k = \big(\omega_0-\omega_k\big)/2$ is a detuning between the main oscillator and the environment oscillators. To understand the meaning of $f(t)$ and $g_{k}(t)$ we will discuss the distribution of excitations between
the main oscillator and of the environment.

\subsection{Dynamics of excitations}
The unitary evolution given by the Hamiltonian (\ref{hamiltoniano}) preserves the total excitation number and therefore
the number of excitation in the main oscillator plus the excitations in the $N$ other will be conserved throughout evolution. In order to discuss the dynamics of excitations we will consider the initial state of the main oscillator in a coherent state $|\alpha(0)\rangle$ and each given oscillator of the environment in the vacuum state $|0_k(0)\rangle$. The initial excitation number average of the global system in this manner is given by
\begin{eqnarray}
\bar{N}_{exc}(0) &=& \langle\psi(0)|a^\dagger a|\psi(0)\rangle.\nonumber \\
\end{eqnarray}
For any time later there is some transfer of excitation between the main oscillator and the environment and therefore the excitation number average of
the global system is given by
\begin{eqnarray}\label{n0}
\bar{N}_{exc}(t) &=& \langle\psi(t)|a^\dagger a+\sum_k b_k^\dagger b_k|\psi(t)\rangle. \nonumber \\
\end{eqnarray}
Using (\ref{alpha_01}) and (\ref{lambda_01}) it can be shown that
\begin{eqnarray}
\langle\psi(t)|a^\dag a|\psi(t)\rangle &=& \langle\psi(0)|a^\dag a|\psi(0)\rangle\Xi(t), \label{exc_princ}\\
\nonumber\\
\sum_k\langle\psi(t)|b^{\dag}_{k}b_k|\psi(t)\rangle &=&
\label{exc_banho} \langle\psi(0)|a^\dag a|\psi(0)\rangle\Theta(t),
\end{eqnarray}
with $\Xi(t)=|f(t)|^2$ and $\Theta(t)=\sum_{k}|g_{k}(t)|^2$. Using now (\ref{exc_princ}) and (\ref{exc_banho}) in (\ref{n0}) we have
\begin{equation}
\frac{\langle\psi(0)|\Big(a^\dag a + \sum_kb_{k}^{\dag}b_k\Big)|\psi(0)\rangle}{\langle\psi(0)|a^\dag a|\psi(0)\rangle} =
\Xi(t)+\Theta(t).\label{prob_exc}
\end{equation}
Recalling that the initial state of the system is given by $|\psi(0)\rangle=|\alpha(0)\rangle\prod_k|0_k\rangle$, consequently
we obtain $\Xi(t)+\Theta(t)=1$ in the right side of Eq. (\ref{prob_exc}). The parameters $\Xi(t)$ and $\Theta(t)$ contain
information about the normalized number of the excitation in each subsystem. In fact, one may say that the system dynamic is completely
given by these parameters and the initial state of the system. To estimate these parameters we need to solve the set of equations given by
(\ref{dif_f_01}) and (\ref{dif_g_01}).

\begin{figure}[th]
\centerline{\psfig{file=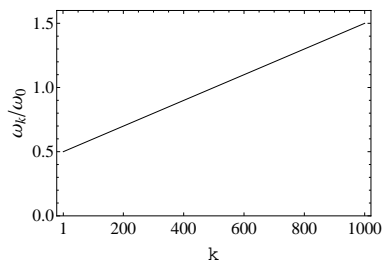,width=5cm}}
\vspace*{8pt}
\caption{Variation of frequency distribution of the environment as a function of the frequency of the main oscillator, $\omega_{k}/\omega_{0}$ (dimensionless). The environment is composed of a set of oscillators with equally spaced frequencies $\omega_{k}$ varying in the range from $0.5\omega_{0}$ to $1.5\omega_{0}$. The $k$ parameter is the $k_{th}$ of the $N$ oscillators of environment with $k = 1, 2, 3, \ldots,1000$.\label{fig2}}
\end{figure}

\begin{figure}[th]
\centerline{\psfig{file=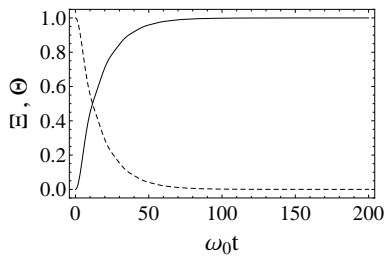,width=5cm}}
\vspace*{8pt}
\caption{Variation of functions $\Xi(t)$ (dashed line) and $\Theta(t)$ (full line) as a function of the (dimensionless) scaled time $\omega_0 t$. For all time $\Xi(t) + \Theta(t) = 1$.\label{fig3}}
\end{figure}

From now on, the system of units is chosen so that $\hbar\omega_0 = 1$, and the values of all physical quantities below are given with
respect to this system. The environment is composed of a set of oscillators with equally spaced frequencies $\omega_{k}$ varying in the range
from $0.5\omega_{0}$ to $1.5\omega_{0}$, see Fig. (\ref{fig2}), and the coupling constant $\gamma_k = \frac{0.1}{\sqrt{N}}$. In the distribution
shown in Fig. (\ref{fig2}) there is at least one oscillation of environment in resonance with the main oscillator ($\omega_{0} = \omega_{k}$).
In Fig. (\ref{fig3}), we show the exchange of excitation between the main oscillator and the environment. As expected, the environment removes all excitations of the main oscillator. Note that, the curve of the main oscillator resembles an exponential decay, like a typical Markovian behavior, where the system of interest loses  excitation to the environment without recovering them. The two curves are complementary so that their sum is equal to $1$, agreeing thereby with the conservation of number of excitations.

\subsection{Excitations of the environment}
\begin{figure}[th]
\centerline{\psfig{file=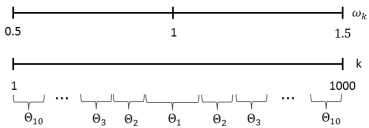,width=5cm}}
\vspace*{8pt}
\caption{The figure illustrate as was divided the environment. It was divided into $10$ partitions with $100$ oscillators each. In
each partition half of the oscillators has a higher frequency, while the other half has a low
frequency in comparison to frequency of the main oscillator.\label{fig4}}
\end{figure}

\begin{figure}[th]
\centerline{\psfig{file=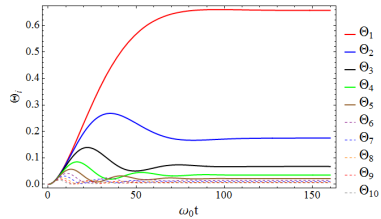,width=5cm}}
\vspace*{8pt}
\caption{Variation of functions $\Theta_i$ of the environment, as a function of the (dimensionless) scaled time $\omega_{0}t$. Each partition $i$ contains $100$ oscillators where the half of the oscillators has a higher frequency, while the other half has a low
frequency in comparison to frequency of the main oscillator. The figure describe $\Theta_i$ with $i = 1, 2, \ldots, 10$ partitions of the environment from top to bottom respectively.\label{fig5}}
\end{figure}

In dynamics described by Fig. (\ref{fig3}) it shows that after some time the environment removes all excitations of the main oscillator. So, we hope that, the interaction between the main oscillator and the environment creates correlations among the environment oscillators. We will now discuss the excitations in the environment and to do so we divide the environment into multiple partitions and after that we will see how the excitations are distributed among these partitions. We divide the environment into $10$ partitions with $100$ oscillators each. In each
partition half of the oscillators has a higher frequency, while the other half has a low frequency in comparison to frequency of the main oscillator.
Figure (\ref{fig4}) shows how this partition was made and Fig. (\ref{fig5}) shows how the excitations evolves in these partitions. Initially there are some oscillations that occur because during this time the main excitation oscillator exchange with the environment. Therefore, while the excitations are being distributed to the environment oscillators, these ones exchange excitations with each other through the main oscillator. After that, all excitations are transferred to the environment, the system is stabilized and there are not exchange of excitations. Therefore, the average number of excitations for
each partition is constant.

\section{Concurrence among the environment oscillators}
Until now we have studied the dynamics of the system. It was obtained the equations of motion and calculated the $\Xi(t)$ e $\Theta(t)$ functions.
We discussed how the excitations are transferred from the main oscillator to the environment and how they are distributed in the environment
after that the system is stabilized. In dynamics described by Fig. (\ref{fig3}) it shows that after some time the environment removes all excitations of the main oscillator and Fig. (\ref{fig5}) shows how the excitations evolves in the partitions of the environment. Since that all the excitations are contained in the environment we hope that there is some entanglement between the environment partitions. Thus, we can study the dynamics of entanglement between the environment partitions. In order to investigate the dynamics of entangled states we will consider the initial state of the system given by
\begin{equation}
|\Psi(0)\rangle=\mathcal{N}\bigl(a|\alpha(0)\rangle+b|\beta(0)\rangle\bigr)\prod_k|0_{k}(t)\rangle
\label{estado_inicial_fatorado2}
\end{equation}
with the initial state of the main oscillator in a coherent superposition of coherent states and each oscillator of the environment in a
vacuum state. The parameters $a$ and $b$ are arbitrary complex number and $\mathcal{N}$ is a normalization constant given by
\begin{equation}
\mathcal{N}^{-1}=\sqrt{|a|^2+|b|^2+a^*b\langle\alpha(0)|\beta(0)\rangle+ab^*\langle\beta(0)|\alpha(0)\rangle}.
\end{equation}
The time evolution of this initial state (\ref{estado_inicial_fatorado2}) is given by
\begin{equation}
|\Psi(t)\rangle=\mathcal{N}\big(a|\alpha(t)\rangle|\lambda(t)\rangle+b|\beta(t)\rangle|\chi(t)\rangle\big)\label{estado_final_fatorado2}
\end{equation}
with
\begin{eqnarray}
|\lambda(t)\rangle &=& \prod_k|\lambda_k(t)\rangle,\nonumber\\
|\chi(t)\rangle &=& \prod_k|\chi_k(t)\rangle. \label{estadoproduto}
\end{eqnarray}
Due to the linearity of the Schr\"odinger equation, the labels ($\alpha(t)$, $\beta(t)$) will obey Eq. (\ref{dif_f_01}) and
($\lambda_k(t)$, $\chi_k(t)$) will obey Eq. (\ref{dif_g_01}) in the form given by Eqs. (\ref{alpha_01}) and (\ref{lambda_01}). For an arbitrary time
$t$ the system becomes entangled in any bipartition.

\begin{figure}[th]
\centerline{\psfig{file=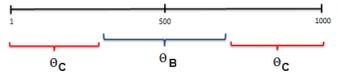,width=5cm}}
\vspace*{8pt}
\caption{The figure illustrate the set of oscillators of the
environment, the partitions $B$, centered at the resonance frequency
of the main oscillator $A$, and the partition $C$ is complementary to the partition $B$.\label{fig6}}
\end{figure}

\begin{figure}[th]
\centerline{\psfig{file=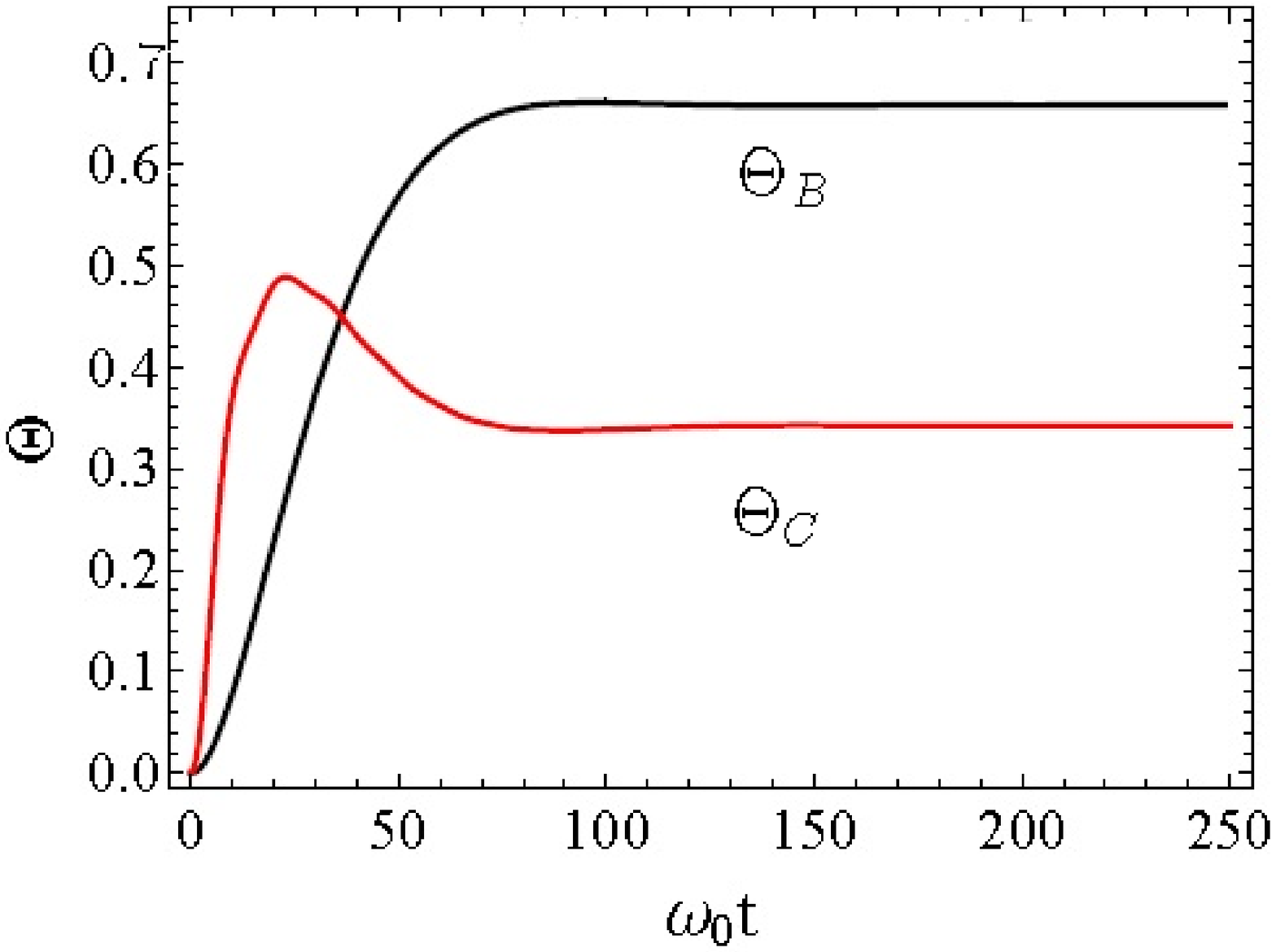,width=5cm}}
\vspace*{8pt}
\caption{The figure describe the variation of functions $\Theta_B$ and $\Theta_C$, as a function of the (dimensionless) scaled time $\omega_{0}t$, with partition $B$ of the environment centered at the resonance frequency of the main oscillator. The partition $B$ contains $100$ oscillators and the partition $C$ contains $900$ oscillators.\label{fig7}}
\end{figure}

\begin{figure}[th]
\centerline{\psfig{file=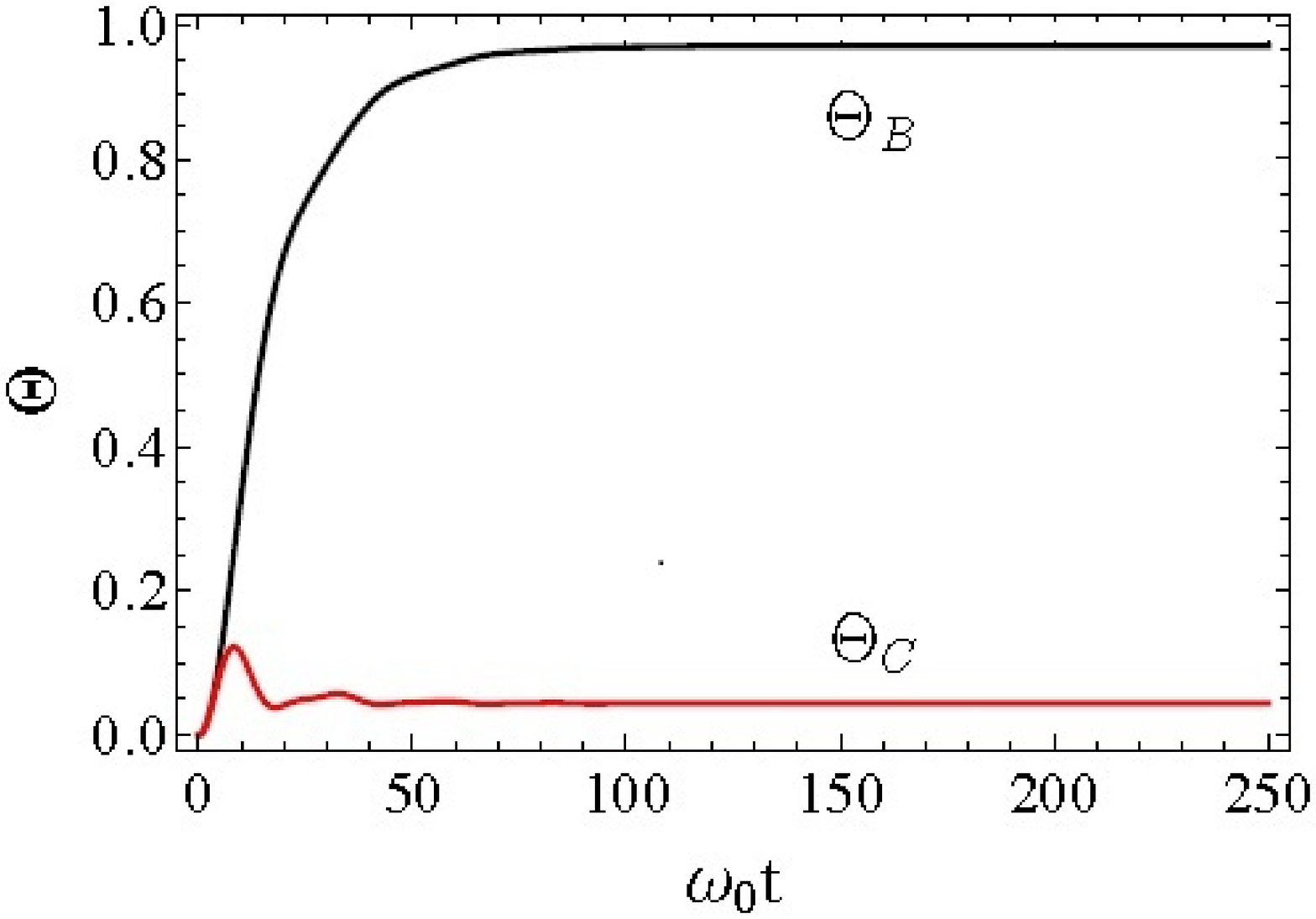,width=5cm}}
\vspace*{8pt}
\caption{The figure describe the variation of functions $\Theta_B$ and $\Theta_C$, as a function of the (dimensionless) scaled time $\omega_{0}t$, with partition $B$ of the environment centered at the resonance frequency of the main oscillator. The partition $B$ contains $500$ oscillators and the partition $C$ contains $500$ oscillators.\label{fig8}}
\end{figure}

\begin{figure}[th]
\centerline{\psfig{file=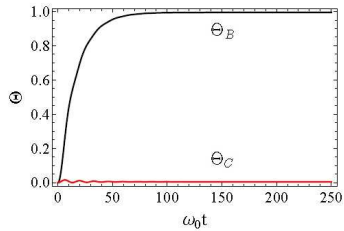,width=5cm}}
\vspace*{8pt}
\caption{The figure describe the variation of functions $\Theta_B$ and $\Theta_C$,as a function of the (dimensionless) scaled time $\omega_{0}t$ , with partition $B$ of the environment centered at the resonance frequency of the main oscillator. The partition $B$ contains $900$ oscillators and the partition $C$ contains $100$ oscillators.\label{fig9}}
\end{figure}

To discuss the distribution of the dynamics of entanglement among the various possible partitions of the system we will calculate the
density matrix of the system given by
\begin{eqnarray}
  \rho_{(s,r)}&=&\mathcal{N}^2\Big\{|a|^2|\alpha(t),\lambda(t)\rangle\langle\alpha(t),\lambda(t)|+|b|^2|\beta(t),\chi(t)\rangle\langle\beta(t),\chi(t)|\nonumber\\
&+& \big[ab^{*}|\alpha(t),\lambda(t)\rangle\langle\beta(t),\chi(t)|+H.c\big]\Big\},\label{mdt}
\end{eqnarray}
where H.c. is the Hermitian conjugate and as described above, the state $|\alpha(t)\rangle$ and $|\beta(t)\rangle$ ($|\lambda(t)\rangle$
and $|\chi(t)\rangle$) are states of the main oscillator (environment). From Eq. (\ref{mdt}) we can easily obtain
the density matrix corresponding to the oscillators of environment. To do this, we will just take the partial trace over the main oscillator. We get
\begin{eqnarray}\label{mdkj}
\rho_{(r)} & = & \mathcal{N}^2\Big\{|a|^2|\lambda(t)\rangle\langle\lambda(t)|+|b|^2|\chi(t)\rangle\langle\chi(t)|
+ \nonumber\\
&& \big[ab^{*}\langle\beta(0)|\alpha(0)\rangle^{\Xi(t)}|\lambda(t)\rangle\langle\chi(t)|+H.c.\big]\Big\},
\end{eqnarray}
where was using $\langle\beta(t)|\alpha(t)\rangle = \langle\beta(0)|\alpha(0)\rangle^{\Xi(t)}$. Writing the state of the environment as a product of two partitions (B and C) (see Eq. (\ref{estadoproduto})), so that:
\begin{eqnarray}\label{estadoproduto2}
|\lambda(t)\rangle &=& |\lambda_B\rangle \otimes|\lambda_C\rangle, \nonumber \\
|\chi(t)\rangle &=& |\chi_B\rangle \otimes|\chi_C\rangle.
\end{eqnarray}
We can now write the reduced density matrix for the environment partitions as:
\begin{eqnarray}
\rho_{(B,C)}&=&\mathcal{N}^2\Big\{|a|^2|\lambda_B,\lambda_C\rangle\langle\lambda_B,\lambda_C|+|b|^2|\chi_B,\chi_C\rangle\langle\chi_B,\chi_C|\nonumber\\
&+& \big[ab^{*}\langle\beta(0)|\alpha(0)\rangle^{\Xi(t)}|\lambda_B,\lambda_C\rangle\langle\chi_B,\chi_C|+H.c.\big]\Big\}.
\label{mdkj2}
\end{eqnarray}
In order to characterize this entanglement we will
rewrite the concurrence used by Wootters to describe entanglement of formation of an arbitrary state of two qubit \cite{wootters98} in
terms of distinguishability of the states of systems. However, we are working with coherent states, which are non-orthogonal states. To obtain the
concurrence in this case we have to write these states in terms of an orthonormal base [see Eq. (\ref{conc_globadem}) in \ref{AppendixA}]. From Eq.(\ref{mdkj2}) the concurrence can be write in the form
\begin{equation}
C_{(B,C)}=2|ab|\mathcal{N}^2|\langle\alpha(0)|\beta(0)\rangle|^{\Xi(t)}\mathcal{D}_B\mathcal{D}_C\label{conc_global}
\end{equation}
where $\mathcal{D}_B$ ($\mathcal{D}_C$) is the distinguishability of the state $|\lambda_B\rangle$ and $|\chi_B\rangle$
($|\lambda_C\rangle$ and $|\chi_C\rangle$) of the partition B (partition C) of the environment. The function $\mathcal{D}_i$ is a reflex of
"distinguishability" between the states $|\mu_i\rangle$ and $|\nu_i\rangle$ and if the overlap between the states
$|\mu_i\rangle$ and $|\nu_i\rangle$ is close to unity the distinctness comes close to zero and the concurrence which is directly proportional
to this tends to disappear \cite{berthold96}. The distinguishability between two coherent states can be written as
\begin{eqnarray}
\mathcal{D}_B &=& \sqrt{1-|\langle\lambda_B|\chi_B\rangle|^2},\label{distsist} \\
\mathcal{D}_C &=& \sqrt{1-|\langle\lambda_C|\chi_C\rangle|^2}.\label{distres}
\end{eqnarray}
From (\ref{exc_princ}) and (\ref{exc_banho})
\begin{eqnarray}
|\langle\lambda_B|\chi_B\rangle|^2 &=& |\langle\alpha(0)|\beta(0)\rangle|^{2\Theta_B}, \label{over1} \\
|\langle\lambda_C|\chi_C\rangle|^2 &=&
|\langle\alpha(0)|\beta(0)\rangle|^{2\Theta_C}. \label{over2}
\end{eqnarray}
In this case, the concurrence is given by
\begin{eqnarray}
C_{B,C} & = & 2|ab|\mathcal{N}^2|\langle\alpha(0)|\beta(0)\rangle|^{\Xi(t)}\sqrt{1-|\langle\alpha(0)|\beta(0)\rangle|^{2\Theta_B}}
\nonumber\\
&& \sqrt{1-|\langle\alpha(0)|\beta(0)\rangle|^{2\Theta_C}}.\label{cbanho}
\end{eqnarray}
Note that, in the Eq.(\ref{cbanho}) there is a term in the expression of concurrence that refers to a subsystem which is outside of the system of interest. The external contribution term involves the $\Xi(t)$ that is directly related to the dynamics of excitations of the main oscillator. According to Eq. (\ref{cbanho}), the concurrence depends on the functions $\Theta_B$ and $\Theta_C$. These functions are related to the dynamics of excitations of the partitions $B$ and $C$ respectively, and therefore, vary with the number of oscillators and the frequency band of each partition. We will analyze the case where the partition $B$ is centered on the frequency of the main oscillator, while the partition $C$ is complementary to the first. Figure (\ref{fig6}) illustrates how this division has been made and the figures (\ref{fig7})-(\ref{fig9}) show how the excitations vary in these partitions.
\\
\\

\begin{figure}[th]
\centerline{\psfig{file=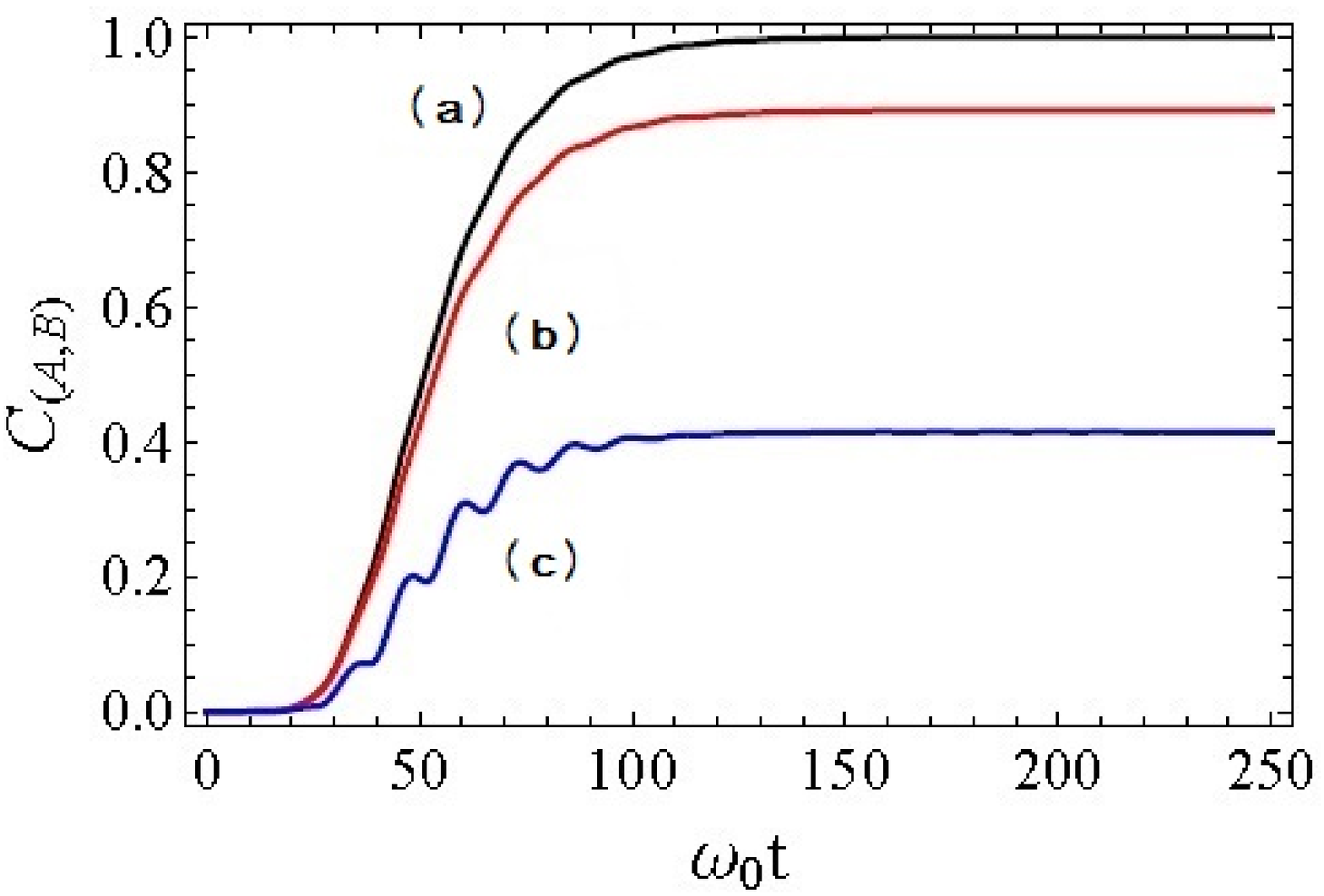,width=5cm}}
\vspace*{8pt}
 \caption{Evolution of the concurrence between the partition $B$ and $C$ as a function of the (dimensionless) scaled time $\omega_{0}t$ to the cases analyzed in figures (\ref{fig7})-(\ref{fig9}) for $\langle\alpha_0|\beta_0\rangle=e^{-18}$ and $b = -a$. In the curve (a) the partition $B$ contains $100$ oscillators, (b) the partition $B$ contains $500$ oscillators and (c) the partition $B$ contains $900$ oscillators.\label{fig10}}
\end{figure}

In Fig. (\ref{fig10}) we plot the evolution of the concurrence to the cases above and it may be observed that in the case that the partition $B$ contains $100$ oscillators the system gets entangled much more than in the case that the partition $B$ contains $900$ oscillators.
To understand this behavior, we will return to concurrence expression. Since we are interested in the situation where all excitation are contained in the environment, $\Xi(T) = 0$, the Eq. (\ref{cbanho}) reduces to the form:
\begin{equation}\label{cbanho2}
C_{B,C}=2|ab|\mathcal{N}^2\sqrt{1-|\langle\alpha(0)|\beta(0)\rangle|^{2\Theta_B}}\sqrt{1-|\langle\alpha(0)|\beta(0)\rangle|^{2\Theta_C}}.
\end{equation}
If $\Theta_B$ or $\Theta_C$ are close to zero, the distinguishability is close to zero, therefore the concurrence tends to zero. Looking at the case where the partition $B$ contains $900$ oscillators, we see through Fig. (\ref{fig9}) that $\Theta_C$ is near zero. So, in this case, the distinguishability of the states of the oscillators partition $C$ is small. Therefore, the system state approaches a separable state. In this way, these two subsystems are little tangled, as can seen in the (c) curve in Fig.(\ref{fig10}). When $\Theta_B$ is close to $\Theta_C$ the distinguishability approaches its maximum value. We see that in the cases studied, the situation where the two functions appear to be closer is shown  in Fig. (\ref{fig7}), where partition $B$ contains $100$ oscillators. As in this situation, the states of each part of the environment are more distinguishable, the entanglement between the two parties becomes greater (see the (a) curve in Fig.(\ref{fig10})). Note that, in accordance with equation (\ref{cbanho2}), in the case that $\Theta_B = \Theta_C = \frac{1}{2}$, the concurrence reaches the value one. That is, if we could separate the environment into two parts, so that the average number of excitations in each of the parts are equal, these two subsystems are maximally entangled. We observed, in this situation, that entanglement was maximum when each subsystem had the same average number of excitations.

Looking at Fig. (\ref {fig10}), we see that the concurrence is constant after some time. After this time, all the excitations are contained in the environment and the main oscillator is in the vacuum state. When this happens, the system dynamics dies. The system is stabilized and there is no exchange of excitations, even among the environment oscillators. Therefore, the functions $\Theta_B$ and $\Theta_C$ remain constant, and consequently the concurrence too. Note that initially the environment oscillator were uncorrelated, there was no entanglement among them. When interacting with the main oscillator, they correlate. So this interaction with the main oscillator creates correlations among the environment oscillators. The superposition of the initial state is transferred to the states of the environment oscillators making them tangled.

\section{Conclusions}
It has been considered here a many-body system consisting of a harmonic oscillator linearly coupled to $N$ others and solved the corresponding dynamical problem analytically. The interaction between the main oscillator and the environment creates correlations among the environment oscillators. Initially the environment oscillators were uncorrelated, there was no entanglement among them. After the interaction, the superposition of the initial state is transferred to the states of the environment creates correlations among the environment oscillators. We have studied the dynamics of entanglement among of the environment oscillators . The entanglement between any two environment partitions was quantified and it showed that in the case that $\Theta_B$ is close to $\Theta_C$ the entanglement between the two parties becomes greater. That is, if we could separate the environment into two parts, so that the average number of excitations in each of the parts are equal, these two subsystems are maximally entangled. We observed, in this situation, that entanglement was maximum when each subsystem had the same average number of excitations.



\appendix
\section{Getting the expression of the concurrence for coherent states}\label{AppendixA}

In this Appendix, we will show the deduction of the concurrence for coherent states used in this paper. To do this, we consider the arbitrary density matrix of a system constituted of two subsystems $\mathcal{A}$ and $\mathcal{B}$ given by:
\begin{equation}
\rho=G\left(p|\alpha,\lambda\rangle\langle\alpha,\lambda|+q|\beta,\chi\rangle\langle\beta,\chi|
+z|\alpha,\lambda\rangle\langle\beta,\chi|+z^*|\beta,\chi\rangle\langle\alpha,\lambda|\right),\label{dgeral}
\end{equation}
where $p$, $q$, z and $G$ are parameter of the system and $\{|\alpha\rangle,|\beta\rangle\}$ ($\{|\lambda\rangle,|\chi\rangle\}$) are coherent states of the subsystem $\mathcal{A}$ ($\mathcal{B}$).
In order to rewrite the concurrence used by Wootters to describe entanglement of formation of an arbitrary state of two qubit in terms of coherent states,
which are non-orthonormal states, we have to write these coherent states in terms of an orthonormal base. Order to do this, we will consider the generic states \cite{tales12}
\begin{eqnarray}
|f\rangle &=& \cos\mu|\alpha(t)\rangle  + e^{i\theta}\sin\mu|\beta(t)\rangle,\nonumber \\
|g\rangle &=& -\sin\mu|\alpha(t)\rangle + e^{i\theta}\cos\mu|\beta(t)\rangle,
\end{eqnarray}
with $\theta$ real. For that the states above are orthogonal $\mu = \frac{\pi}{4}$ and $e^{i\theta} = \frac{\langle\beta(t)|\alpha(t)\rangle}{|\langle\beta(t)|\alpha(t)\rangle|}$. Normalizing these states
\begin{eqnarray}
|1\rangle &=& \frac{|f\rangle}{\sqrt{\langle f|f\rangle}}=\frac{1}{2S_{+}}\left(|\alpha(t)\rangle+e^{i\theta}|\beta(t)\rangle\right), \nonumber \\
|0\rangle &=& \frac{|g\rangle}{\sqrt{\langle g|g\rangle}}=\frac{1}{2S_{-}}\left(-|\alpha(t)\rangle+e^{i\theta}|\beta(t)\rangle\right),
\end{eqnarray}
with $S_{\pm}=\sqrt{\frac{1\pm|\langle\alpha(t)|\beta(t)\rangle|}{2}}$. In a similar procedure we can write another pair of orthonormal states taking into account the environment variables, ie:
\begin{eqnarray}
  |\uparrow\rangle &=& \frac{1}{2S'_{+}}\left(|\lambda(t)\rangle+e^{i\theta'}|\chi(t)\rangle\right), \nonumber \\
  |\downarrow\rangle &=& \frac{1}{2S'_{-}}\left(-|\lambda(t)\rangle+e^{i\theta'}|\chi(t)\rangle\right),
\end{eqnarray}
with $S'_{\pm}=\sqrt{\frac{1\pm|\langle\lambda(t)|\chi(t)\rangle|}{2}}$ and $e^{i\theta'} = \frac{\langle\chi(t)|\lambda(t)\rangle}{|\langle\chi(t)|\lambda(t)\rangle|}.$
We can now write the coherent states in terms of these orthonormal states. So we obtain that
\begin{eqnarray}
|\alpha(t)\rangle &=& S_{+}|1\rangle-S_{-}|0\rangle, \nonumber \\
|\beta(t)\rangle &=& e^{-i\theta}\left(S_{+}|1\rangle+S_{-}|0\rangle\right),\nonumber \\
|\lambda(t)\rangle &=& S'_{+}|\uparrow\rangle-S'_{-}|\downarrow\rangle, \nonumber \\
|\chi(t)\rangle &=& e^{-i\theta'}\left(S'_{+}|\uparrow\rangle+S'_{-}|\downarrow\rangle\right)\label{bcoer}.
\end{eqnarray}
The states $\{|1\rangle,|0\rangle\}$ and $\{|\uparrow\rangle,|\downarrow\rangle\}$ form a base for our system (qubits). Therefore, for any instant in time, the system can be written as a pair of qubit. After this change of base, we can calculate the concurrence between the two subsystems.

Now, using the coherent states given by (\ref{bcoer}) in (\ref{dgeral}) we obtain the density matrix
\begin{eqnarray}
\rho &=& G\left(S_+^2S_+^{'2}r|1\uparrow\rangle\langle1\uparrow|+S_+^2S_+^{'}S_-^{'}u|1\uparrow\rangle\langle1\downarrow|+S_+S_-S_+^{'2}u|1\uparrow\rangle\langle0\uparrow|\right.\nonumber \\
&\,&+ S_+S_-S_+^{'}S_-^{'}r|1\uparrow\rangle\langle0\downarrow|+S_+^{2}S_+^{'}S_-^{'}u^*|1\downarrow\rangle\langle1\uparrow|+S_+^{2}S_-^{'2}v|1\downarrow\rangle\langle1\downarrow|\nonumber\\
&\,&+ S_+S_-S_+^{'}S_-^{'}v|1\downarrow\rangle\langle0\uparrow|+S_+S_-S_-^{'2}u^*|1\downarrow\rangle\langle0\downarrow|+S_+S_-S_+^{'2}u^*|0\uparrow\rangle\langle1\uparrow|\nonumber\\
&\,&+ S_+S_-S_+^{'}S_-^{'}v|0\uparrow\rangle\langle1\downarrow|+S_-^{2}S_+^{'2}v|0\uparrow\rangle\langle0\uparrow|+S_-^{2}S_+^{'}S_-^{'}u^*|0\uparrow\rangle\langle0\downarrow|\nonumber\\
&\,&+ S_+S_-S_+^{'}S_-^{'}r|0\downarrow\rangle\langle1\uparrow|+S_+S_-S_-^{'2}u|0\downarrow\rangle\langle1\downarrow|+S_-^{2}S_+^{'}S_-^{'}u|0\downarrow\rangle\langle0\uparrow|\nonumber\\
&\,&+\left. S_-^{2}S_-^{'2}r|0\downarrow\rangle\langle0\downarrow|\right),
\end{eqnarray}
with
\begin{eqnarray}
r &=& p+q+ze^{i(\theta+\theta')}+z^*e^{-i(\theta+\theta')};\nonumber \\
u &=& -p+q+ze^{i(\theta+\theta')}-z^*e^{-i(\theta+\theta')};\nonumber\\
v &=& p+q-ze^{i(\theta+\theta')}-z^*e^{-i(\theta+\theta')}.
\end{eqnarray}
In the matrix form is given by:
\begin{equation}
\rho=G\left(
\begin{array}{cccc}
 S_+^2S_+^{'2}r & S_+^2S_+^{'}S_-^{'}u & S_+S_-S_+^{'2}u & S_+S_-S_+^{'}S_-^{'}r \\
 S_+^{2}S_+^{'}S_-^{'}u^* & S_+^{2}S_-^{'2}v & S_+S_-S_+^{'}S_-^{'}v & S_+S_-S_-^{'2}u^* \\
 S_+S_-S_+^{'2}u^* & S_+S_-S_+^{'}S_-^{'}v & S_-^{2}S_+^{'2}v & S_-^{2}S_+^{'}S_-^{'}u^* \\
 S_+S_-S_+^{'}S_-^{'}r & S_+S_-S_-^{'2}u & S_-^{2}S_+^{'}S_-^{'}u & S_-^{2}S_-^{'2}r \\
\end{array}
\right).
\end{equation}

Having defined the base of qubits for our system, we can now rewrite the concurrence used by Wootters in terms of coherent states. The concurrence to describe entanglement of formation of an arbitrary state of two qubit can be defined as $C=max(0,l_1-l_2-l_3-l_4)$\cite{wootters98}, where $l_i$ are the eigenvalues, in decreasing order, of the Hermitian matrix $R=\sqrt{\sqrt{\rho}\tilde{\rho}\sqrt{\rho}}$. Here $\tilde{\rho}$ is obtain using the \emph{spin-flip} transformation
\begin{equation}\label{s-flip}
\tilde{\rho}=\sigma_y\rho^*\sigma_y=(\sigma_{1y}\otimes\sigma_{2y})\rho^*(\sigma_{1y}\otimes\sigma_{2y}),
\end{equation}
where $\rho^*$ is the complex conjugate of $\rho$ and $\sigma_y = \sigma_{1y}\otimes\sigma_{2y}$ is the matrix's Pauli of the correspondent system. The concurrence can be write as $C=max(0,2l_{max}-Tr\{R\})$ where $Tr\{R\}=Tr\{\sqrt{\sqrt{\rho}\tilde{\rho}\sqrt{\rho}}\}$. Using now $\sqrt{\rho} = U\sqrt{\rho_D}U^\dagger$, where $\rho_D$ is the diagonal form of $\rho$ and $U$ is the matrix that diagonalize $\rho$, we get:
\begin{eqnarray}
Tr \{R\} &=& Tr\{\sqrt{\sqrt{\rho}\tilde{\rho}\sqrt{\rho}}\}\nonumber \\
    &=& Tr\{U\sqrt{\sqrt{\rho_D}}U^\dagger\sqrt{\tilde{\rho}}U\sqrt{\sqrt{\rho_D}}U^\dagger\}\nonumber \\
    &=& Tr\{\sqrt{\rho_D}U^\dagger\sqrt{\tilde{\rho}}U\}\nonumber \\
    &=& Tr\{\sqrt{\rho\tilde{\rho}}\} = Tr\{\sqrt{M}\},
 \end{eqnarray}
where $M = \rho\tilde{\rho}$. As $Tr \{R\}=\sum_i l_i$, where $l_i$ are the eigenvalues of $R$ and $Tr\{\sqrt{M}\}=\sum_i\sqrt{m_i}$, where $m_i$ are the eigenvalues of $M$, we obtain $l_i=\sqrt{m_i}$. Therefore to find the expression of concurrence we have to obtain $M$ and get its eigenvalues. To obtain $M$, first we need to calculate $\tilde{\rho}$ using the operation (\ref{s-flip}).

Now, write the matrix $\sigma_y$ in the base given by (\ref{bcoer}) we have
\begin{eqnarray}
  \sigma_y &=& (-i|1\rangle\langle0|+i|0\rangle\langle1|)\otimes(-i|\uparrow\rangle\langle\downarrow|+i|\downarrow\rangle\langle\uparrow|)\nonumber \\
   &=& - |1\uparrow\rangle\langle0\downarrow|-|0\downarrow\rangle\langle1\uparrow|+|1\downarrow\rangle\langle0\uparrow|+|0\uparrow\rangle\langle1\downarrow|.
\end{eqnarray}
Using $\sigma_y$ in the operation describe in (\ref{s-flip}), we get
\begin{eqnarray}
  \tilde{\rho} &=& G\left(S_+^{2}S_+^{'2}r |0\downarrow\rangle\langle0\downarrow|+S_+^{2}S_-^{'2}v|0\uparrow\rangle\langle0\uparrow|+S_-^{2}S_+^{'2}v|1\downarrow\rangle\langle1\downarrow|\right.\nonumber\\
   &\,& + S_-^{2}S_-^{'2}r|1\uparrow\rangle\langle1\uparrow| +\{-S_+^{2}S_+^{'}S_-^{'}u|0\uparrow\rangle\langle0\downarrow|-S_+S_-S_+^{'2}u|1\downarrow\rangle\langle0\downarrow| \nonumber\\
   &\,& + S_+S_-S_+^{'}S_-^{'}v|1\downarrow\rangle\langle0\uparrow|+S_+S_-S_+^{'}S_-^{'}r|1\uparrow\rangle\langle0\downarrow| - S_+S_-S_-^{'2}u^*|1\uparrow\rangle\langle0\uparrow|\nonumber\\
   &\,&-S_-^{2}S_+^{'}S_-^{'}u^*|1\uparrow\rangle\langle1\downarrow| +c.h\}),
\end{eqnarray}
and in the matrix form
 \begin{equation}
 \tilde{\rho}=G\left(
\begin{array}{cccc}
 S_-^{2}S_-^{'2}r & -S_-^2S_+^{'}S_-^{'}u^* & -S_+S_-S_-^{'2}u^* & S_+S_-S_+^{'}S_-^{'}r \\
 -S_-^{2}S_+^{'}S_-^{'}u & S_-^{2}S_+^{'2}v & S_+S_-S_+^{'}S_-^{'}v & -S_+S_-S_+^{'2}u \\
 -S_+S_-S_-^{'2}u & S_+S_-S_+^{'}S_-^{'}v & S_+^{2}S_-^{'2}v & -S_+^{2}S_+^{'}S_-^{'}u \\
 S_+S_-S_+^{'}S_-^{'}r & -S_+S_-S_+^{'2}u^* & -S_+^{2}S_+^{'}S_-^{'}u^* & S_+^2S_+^{'2}r \\
\end{array}
\right).
\end{equation}

Now we can obtain $M=\rho\tilde{\rho}$ and calculate its eigenvalues
\begin{eqnarray}
m_1 = 2{G^2}S_+^{2}S_-^{2}S_+^{'2}S_-^{'2}&\{ &r^2+v^2-u^2-u^{*2}\nonumber \\
&-& \sqrt{(-r^2-v^2+u^2+u^{*2})-(4r^2v^2-8r|u|^2v-4|b|^4)}\}; \nonumber\\
 m_2 = 2{G^2}S_+^{2}S_-^{2}S_+^{'2}S_-^{'2}&\{ &r^2+v^2-u^2-u^{*2}\nonumber \\
&+& \sqrt{(-r^2-v^2+u^2+u^{*2})-(4r^2v^2-8r|u|^2v-4|b|^4)}\}. \nonumber\\ \label{m1m2}
\end{eqnarray}
Using the values of the \emph{r}, \emph{u} and \emph{v} in (\ref{m1m2}) we have
\begin{eqnarray}
  m_1 &=& 16G^2S_+^{2}S_-^{2}S_+^{'2}S_-^{'2}(|z|-\sqrt{pq})^2; \nonumber \\
  m_2 &=& 16G^2S_+^{2}S_-^{2}S_+^{'2}S_-^{'2}(|z|+\sqrt{pq})^2.
\end{eqnarray}

We can now obtain the eigenvalues of $R$:
\begin{eqnarray}
  l_1 &=& 4GS_+S_-S_+^{'}S_-^{'}(\sqrt{pq}-|z|), \nonumber \\
  l_2 &=&  4GS_+S_-S_+^{'}S_-^{'}(\sqrt{pq}+|z|).
\end{eqnarray}
To in the case studied here $\sqrt{pq}>|z|$ the concurrence is given by
\begin{equation}
  C=l_2-l_1=8GS_+S_-S_+^{'}S_-^{'}|z|.\label{concurrence2}
\end{equation}

Note that we start from a general density matrix (\ref{dgeral}) of a system constituted of two subsystems formed by coherent states. To the case studied here we can write
\begin{equation}
  G=\mathcal{N}^2, \qquad p=|a|^2, \qquad q=|b|^2, \qquad z=ab^{*}\langle\beta(0)|\alpha(0)\rangle^{\Xi(t)}
\end{equation}
and use in the equation of concurrence give by (\ref{concurrence2}), we obtain
\begin{equation}
C_{(B,C)}=2|ab|\mathcal{N}^2|\langle\alpha(0)|\beta(0)\rangle|^{\Xi(t)}
\sqrt{1-|\langle\lambda_B|\chi_B\rangle|^2}\sqrt{1-|\langle\lambda_C|\chi_C\rangle|^2}.\label{conc_globadem}
\end{equation}
%
%
%
%
%
%

\end{document}